\documentclass[twocolumn,showpacs,preprintnumbers,amsmath,amssymb,prb]{revtex4}
\newcommand{\be}{\begin{equation}}
\newcommand{\ee}{\end{equation}}
\newcommand{\bea}{\begin{eqnarray}}
\newcommand{\eea}{\end{eqnarray}}
\usepackage{graphicx}
\usepackage{float,afterpage}
\usepackage{dcolumn}
\usepackage{bm}
\begin{document}
\title
{Schr\"odinger equation for the one-particle density matrix of 
thermal systems: An alternative formulation of Bose-Einstein condensation}
\author{Subodha Mishra and Peter Pfeifer}
\affiliation {Department of Physics \& Astronomy,
 University of Missouri, Columbia, MO 65211, USA}
\date{\today}
\begin{abstract}
We formulate a linear Schr\"odinger equation with the 
temperature-dependent 
potential for the one-particle density matrix and obtain the 
condensation temperature of the Bose-Einstein condensate from a 
bound-state condition for the 
Schr\"odinger equation both with and without the confining trap. The 
results are in very good agreement with 
those of the full statistical physics treatment. This is an 
alternative 
to the Bose-Einstein condensation in the standard ideal Bose gas 
treatment.
\end{abstract}
\maketitle
\section{Introduction}
Based on the Bose-Einstein statistics\cite{bose}, Einstein 
preidicted\cite{ein} that for a gas of non-interacting, massive bosons, 
below a 
critical temperature a finite fraction of the total number of particles 
would occupy the lowest-energy single-particle state forming a condensate. 
This phenomenon of Bose-Einstein condensation (BEC) of dilute
gases has been
extensively studied theoretically\cite{leg} and only  
recently it  has been  
experimentally\cite{and,bra,dav} realized 
for dilute atomic gases of $^{87}$Rb, $^{7}$Li, $^{23}$Na. 
It is known that the  Gross-Pitaevskii equation (GPE)\cite{gross}, a 
non-linear 
Schr\"odinger equation 
for the 
macroscopic wave function of the BEC provides a detailed description of 
its ground state spectrum. Though the GPE has been solved 
variationally\cite{mas,shi}, in this paper, we give a new theoretical 
treatment 
of the  BEC by
incorporating the symmetry of the system with a temperature-dependent
potential and solving a linear temperature dependent Hamiltonian 
variationally.
By incorporating a non-linear
term like $\sim |\psi|^2$ in our Hamiltonian one can
construct a finite temperature extension of the Gross-Pitaevskii 
equation\cite{gross}
for an interacting bose gas.
The present
approach can be extended in many ways including the study of temperature
dependence of the energy of a vortex in a Bose gas and effect of
temperature on the interference of two macroscopic Bose-condensates.
%
%
%
The free-particle many-body Hamiltonian\cite{huang} with the 
quantum 
correction due to the 
Bose/Fermi statistics of the system is given as a function of temperature 
as
\be
H=\sum_i\frac{p_i^2}{2m}+\sum_{i,j,i\ne j} \tilde v(|\vec r_i-\vec 
r_j|) 
\label{h}
\ee
where $\vec p_i$ and $\vec r_i$ are the momentum and position of the 
i$^{th}$ particle, the potential is given by
\be
\tilde v(r)= -k_BT \\ \ln\bigg [{1 \pm exp\bigg ({-2\pi\frac{r^2} 
{\lambda^2}}}\bigg )\bigg ]
\ee
and $\lambda$ is the thermal wave length, 
$\lambda=\sqrt{2\pi\hbar^2/{mk_BT}}$. The plus sign is for bosons, 
and the minus sign is for fermions. The Hamiltonian (1) represents first  
quantum correction to the 
classical gas high temperature expansion and describes the system in the 
range where the inter 
particle separation is much greater than the corresponding thermal 
wavelength, $|\vec r_i -\vec r_j| >> \lambda$. The statistical potential 
$\tilde v(r)$ is 
attractive for bosons and repulsive for fermions.  
The one particle Schr\"odinger equation is given as
\be
{-\hbar^2\over 2\mu}\nabla^2\psi + \tilde v\psi=E\psi
\ee
where $\mu=m/2$ is the reduced mass of the two particle system when the 
Hamiltonian
Eq.(\ref{h}) is written  for $i=2$.
The exact thermal one-particle density matrix\cite{ziff} for bose-systems, 
in position representation is given 
as
\bea
\rho_1 (|\vec r -\vec r'|,T) = F(|\vec r -\vec r'|,T) \nonumber \\
=\frac{1}{\lambda^3}\sum_{j=1}^\infty
j^{-\frac{3}{2}}exp(-\alpha j -\pi r^2/{j\lambda^2})\label{rh1}
\eea
where $\alpha$ has the value $0<\alpha\le 1$.
Our claim is the leading term in the one
particle density matrix\cite{ziff} of a bose-system in the position
representation can be
approximated as  $\rho_1^L (r,T)\approx |\psi (r,T)|^2$ where L denotes 
the leading term.
We replace 1-particle Hamiltonian ${-\hbar^2\over 2\mu}\nabla^2 + \tilde 
v$ by 
\be
H_{eff}:={-\hbar^2\over 2\mu}\nabla^2 + \epsilon\tilde{\tilde v}\label{heff1}
\ee
\be
\tilde{\tilde v}:=-k_BT e^{-\frac{2\pi r^2}{\lambda^2}}\label{int}
\ee
and we have approximated $\tilde v(r)\approx \tilde{\tilde v}(r)$ 
for large T, that is when the interparticle separation is much greater 
than the corresponding thermal wavelength, $r>>\lambda$.  
We have multiplied the potentail with  $\epsilon$ to take into
account the higher order correlation effect, and it will be determined
dynamically such that the eigenvalue of the Hamiltonian is $\le 0$.
We derive the dependence of critical condensation density on temperature 
of 
the system with or without a confining trap.
We state an alternate derivation of the quantum potential 
and
analyse our wavefunction in the limiting case in the Appendix.
\begin{figure}[htb]
\includegraphics[angle=-90,width=0.75\linewidth]
{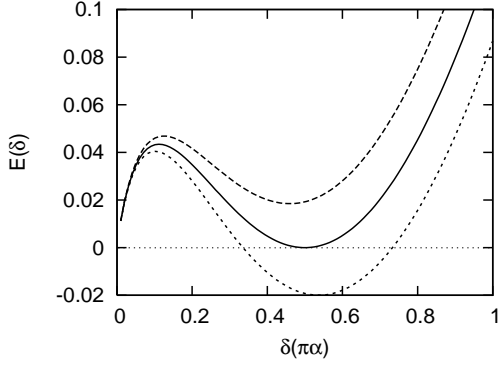}
\caption{\label{fig:bands}
Variation of ground state energy of the bose gas with the parameter
$\delta$  for different values of $\epsilon$. For $\epsilon=3.89711$, the
curve with the solid line shows that the ground state just begins to
form. The long dashed line is for $\epsilon=3.79711$ and the small dashed
line is for $\epsilon=3.99711$.}
\end{figure}
\section{Schr\"odinger's equation in 3-D}
The  time independent 3-D Schr\"odinger equation in spherical polar 
coordinate for the reduced mass $\mu=m/2$ is 
given as 
\bea
{-\hbar^2\over 2\mu}\bigg [\frac{1}{r^2} \frac {\partial}{\partial 
r}r^2\frac{\partial}{\partial r} +\frac{1}{r^2 
sin\theta}\frac{\partial}{\partial \theta}(sin\theta 
\frac{\partial}{\partial \theta})+\frac{1}{r^2 
sin^2\theta}\frac{\partial^2}{\partial \phi^2} \nonumber \\ 
+ \tilde{\tilde v}(r,T)\bigg ]\psi(r,\theta,\phi)=E\psi(r,\theta,\phi)
\eea
where $E$ is the energy eigenvalue corresponding to the wavefunction. We 
write the total wavefunction $\psi(r,\theta,\phi)$, where T is a parameter 
in the Hamiltonian interpreted as temperature, 
as a product of the radial and the angular parts. 
\be
\psi(r,\theta,\phi)=R_{n,l}(r)Y_{l,m}(\theta,\phi)
\ee
Substituting the above $\psi(r,\theta,\phi)$ in the 3-D Schr\"odinger 
equation, we get the equation for $R_{n,l}(r)$ as
\be
\frac{1}{r^2} \frac {\partial}{\partial r}r^2\frac {\partial R}{\partial 
r}+\frac{2\mu}{\hbar^2}[E-\tilde{\tilde v}(r,T)-{l(l+1)\hbar^2\over 2\mu 
r^2}]R(r)=0
\ee
Using $R(r)=\frac{\chi(r)}{r}$,
we get the standard 1-D radial Schr\"odinger equation as
\be
-\frac{\hbar^2}{2\mu}\frac{\partial^2 \chi}{\partial r^2} +[\tilde{\tilde 
v}(r,T)
+\frac{l(l+1)\hbar^2}{2\mu r^2}]\chi=E\chi\label{1d} \\ \nonumber
\ee
\section{BEC without harmonic trap}
\subsection{Variational calculation}
First we study the simple case of the BEC without any confining trap.
For $l=0$,
the  1D Hamiltonian is given from Eq.(\ref{1d}) as 
\be H_{eff}=\frac{-\hbar^2}{2\mu}\frac{\partial^2 }{\partial r^2}
-\epsilon k_BTe^{(-\frac{2\pi r^2}{\lambda^2})}\label{heff}
\ee
We choose the trial wave function $\chi(r)$ as
\be
\chi(r,T)=Are^{-\delta r^2}\label{trial}
\ee
where $\delta$ is a variational parameter which will depend on 
temperature T and $A$ is normalisation 
constant given as 
\be
A=(\frac{2^7 \delta^3}{\pi})^{1\over 4}\label{norm}
\ee
Now we calculate the expectation value of the Hamiltonian $H$ given 
by Eq.(\ref{heff}), which gives us the energy
\be
E=<\chi(r,T)|H|\chi(r,T)>
\ee
We get energy $E$ as 
\be
E=k_B T \bigg [\frac{3\delta}{2\pi\alpha}
- {\epsilon }\bigg ( \frac{1}{1+\pi\alpha
 \delta^{-1}}\bigg )^{(3/2)}\bigg ]\label{energy}
\ee
where $\alpha=\frac{m k_BT}{2\pi\hbar^2}$ and $\delta/\alpha$ is 
dimensionless.
We note, even if temperature  appears in a complicated way 
in the interaction term in Eq.(\ref{int}), the effect of T in the energy 
expression [Eq.(\ref{energy})] is 
multiplicative. 
Now we do the minimization of the energy given by the Eq.(\ref{energy}) with respect 
to $\delta$, which is $\frac{\partial E(\delta,\epsilon)}{\partial 
\delta}=0$.
We get
\be
\frac{\delta}{\pi\alpha} -\epsilon^{2/5} (\frac{\delta}{\pi\alpha})^{1/5} 
+1=0
\ee
Since the bound state starts to appear when the total energy $E\rightarrow 
0^-$ 
and 
the above minimum condition is satisfied, we  therefore solve the 
two  simultaneous equations for the two unknowns $\delta$ and 
$\epsilon$ as 
\bea
E(\epsilon,\delta)=0 \\
\frac{\partial E(\epsilon,\delta)}{\partial \delta}=0
\eea
and we get $\epsilon=\epsilon_c=3.89711$ and 
$\delta=\delta_c=0.5\pi\alpha$. In the next 
section we will see that this set of values of $\epsilon,\delta$ gives 
the right relation  between density and temperature. 
The plot of $E(\delta)$ as a function of  $\delta$ for different values 
of $\epsilon$ is given in Fig.1. This shows for 
$\epsilon=\epsilon_c=3.89711$, 
$\delta=\delta_c=0.5\pi\alpha$ the bound state just begins to form 
which is  onset of BEC(shown as solidline). 
\subsection{The Density-Temperature Relation} 
We now calculate the density of an uniform bose gas in a volume $v$ at 
temperature T at 
which the condensation occurs. We have  effectively one particle in 
volume $v$. We use the standard definition of  volume as $v=<r^3>$ ie, 
\bea
v=\bigg ({\frac{\int r^3 |\psi(r)|^2 d^3r}{\int|\psi(r)|^2 
d^3r}}\bigg )=\bigg ({\frac{\int r^3 |\chi(r)|^2 dr}{\int  |\chi(r)|^2 
dr}}\bigg ) \nonumber \\
=(\frac{2}{\pi\delta^3})^{\frac{1}{2}}\label{v2} 
\eea
we get 
\bea
n_c=\frac{1}{v}
=2.46740\bigg ({\frac{mk_BT_c}{2\pi\hbar^2}}\bigg 
)^{\frac{3}{2}}\label{nc2}
\eea
We see from Eq.(\ref{nc2}) that we reproduce the right 
relation 
between condensation density $n_c$ of the bose gas and the temperature T 
at which the 
condensation occurs from an effective one particle picture.The 
exact relation has the constant\cite{huang} 
$g_{3/2}(1)=\zeta(3/2)=2.612$, where $\zeta(x)$ is the Riemann zeta 
function of $x$.  
The density-temperature relation is the most important result of our 
paper. Since ours is a variational one, the coefficient is little off 
from the exact value. This shows that our 
formalism reproduces the relation which is known otherwise for an ideal 
bose gas\cite{huang}.
\section{BEC in the harmonic trap}
Now we study the BEC in an isotropic 
harmonic trap.
The Hamiltonian for the quantum bose gas in the trap is given as
\be
H=\frac{-\hbar^2}{2\mu}\frac{\partial^2 }{\partial r^2}
-\epsilon k_BTe^{(-\frac{2\pi r^2}{\lambda^2})}
+\frac{1}{2}\mu\omega^2 r^2
\ee
We have added the extra confining potential with frequency $\omega$ to our 
original
Hamiltonian [Eq.(\ref{heff})] and we want to study its effect on the 
BEC.
We use our trial wave function given by Eq.(\ref{trial}) with the usual
normalization constant A given by Eq.(\ref{norm}).
We calculate the energy E and minimize it to get one more equation
to solve for the two parmeters $\epsilon$ and $\delta$.
The two equations are given as 
\be
E=k_B T \bigg [\frac{3\delta}{2\pi\alpha}
- \frac{\epsilon}{(1+\pi\alpha
 \delta^{-1})^{\frac{3}{2}}} +{\frac{3{\sqrt 2}}
{8}} (\frac{\hbar\omega}{k_BT})^2({\frac{\pi\alpha}{\delta}})\bigg ]
\ee
\be
\frac{\partial E}{\partial\delta}=k_B T \bigg [\frac{3}{2\pi\alpha}
- \frac{\epsilon 3(\pi\alpha/\delta)^2} 
{2\pi\alpha(1+\pi\alpha
 \delta^{-1} )^{\frac{5}{2}}} - {\frac{3{\sqrt 2}}{8\pi\alpha}} 
(\frac{\hbar\omega}{k_BT})^2({\frac{\pi\alpha}{\delta}})^2\bigg ]
\ee
Now we use this two equaion with $E=0$ and $\frac{\partial
E}{\partial\delta}=0$  to solve simultaneously for
the two unknowns $\epsilon$ and  $\delta$ at different values of 
temperature T while keeping the value of frequency $\omega$ constant. 
Since now we can not solve for $\delta$ in a closed analytic form as a 
function of T, we do the calculation numerically.
For each given value of temperature T, we solve for $\epsilon$ and 
$\delta$. Then using Eq.(\ref{v2}),we calculate volume $v$ and hence 
condensation density 
$n_c$ for each value of $\delta$ for a set of  values of T. 
We show in Fig.2 the numerical data obtained for
$n_c$ vs T and the fitting  of the data with a function 0.354$T^{3/2}$,
where we have taken $\hbar\omega/k_B=1$. The fitting  is amazingly 
good. From the ideal bose-gas treatment the value of the coefficient 
is 0.167 instead of 0.354. Since ours is a variational one, the 
coefficient is overestimated by a factor of 2. Otherwise our 
formulation
correctly reproduces the density-temperature relation for the bose gas 
confined in a harmonic trap. 
Since we know that  size\cite{peth} of the thermal cloud is given by 
$r\sim
(\frac{k_BT}{m\omega^2})^{\frac{1}{2}}$, using this in Eq.(\ref{nc2}) 
i.e, N$\sim v\times T^{3/2}$
, we get the number
of particles N vary as $T^3$.
This is the relation one gets if one  uses other methods\cite{peth}
for the  calculation. This shows that the single particle representation
of the BEC is very good.
\begin{figure}[htb]
\includegraphics[angle=-90,width=0.75\linewidth]
{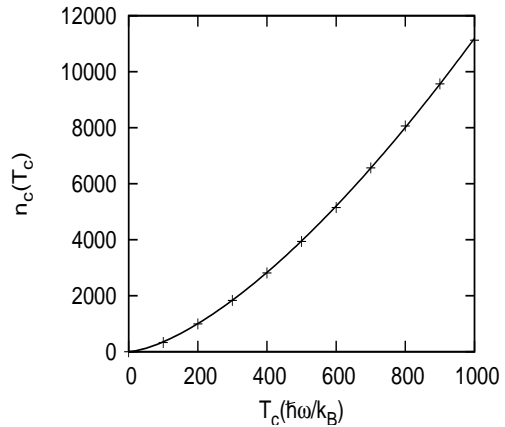}
\caption{\label{fig:bandssss}
Variation of condensation density with temperature. The +'s are the
numerically calculated data and the solid line is a fit function given as
0.354$ T^{3/2}$.}
\end{figure}

\section{Relation of our Trial Wavefunction to the Exact One-particle 
Density Matrix}
For an ideal Bose-Einstein gas it has been shown\cite{ziff} that the one 
particle density matrix is given by Eq.(\ref{rh1}). 
Since $0<\alpha\le 1$, we can see that the leading term in the above 
density matrix with $j=1$ 
has the dependance on $r$ and T as 
\be
\rho_1^L(r)\sim \frac{1}{\lambda^3}e^{-\frac{\pi 
r^2}{\lambda^2}}\label{rho1}
\ee
We can also calculate the one particle density matrix from our trial 
wave function Eq.(\ref{trial}) as 
\be
\rho^{our}_1(r)=\psi^*(r)\psi(r)=\frac{4\pi}{\lambda^3} 
e^{-\frac{\pi r^2}{\lambda^2}}\label{rho2}
\ee
Comparing Eq.(\ref{rho1}) and Eq(\ref{rho2}) we see that the functional 
dependence on $r$ 
and $T$ of both the density matrices are same. Infact Eq.(\ref{rho2}) is 
the 
density matrix\cite{feynman} for a three-dimensional system with $N$ 
particles. So it justifies the one 
particle description of the BEC in our formalism by taking a temperature 
dependent pure state.
\section{Conclusion}
In conclusion, we have shown that it is possible to get Bose-Einstein 
condensation  with or without the harmonic trap 
by considering a Hamiltonian with symmetry dependent potential and a 
temperature dependent pure state. We reproduce the right relation between 
condensation density and temperature at the critical point. This is an 
alternative way 
to the Bose-Einstein condensation in standard ideal bose gas treatment.
By incorporating a non-linear
term like $\sim |\psi|^2$ in the linear Schr\"odinger equation one can
construct a finite temperature extension of the Gross-Pitaevskii
equation\cite{gross}.
The present 
approach can be applied to study the temperature 
dependence of the energy of a vortex in a Bose gas and effect of 
temperature on the interference of two macroscopic Bose-condensates.
\appendix
\section{Derivation of the interaction potential from the distribution 
function}
For a pair of non-interacting particles in a volume $v$ the canonical 
partition function 
is given as 
\be
Z(v,T)\approx {\frac{1}{2}}{(\frac{v}{\lambda^3})^2}
\ee
where $\lambda$ is the thermal de-Broglie wave length defined before.
It can be shown\cite{pathria} that the probability distribution for the 
separation $r$ is given as 
\be
P_{12}=\frac{1}{v^2}\bigg [1\pm exp(-2\pi r^2/{\lambda^2})\bigg ]
\ee
Now we take the logarithm  of the above equation and get
\be
\ln P_{12}=\ln\bigg [1\pm exp(-2\pi r^2/{\lambda^2})\bigg ]
\ee
where +(-) sign is for boson(fermion).
In writting the  above equation we have dropeed the constant term 
independent of $r$. We multiply by $-k_B$ to 
find a quantity which is a kind of entropy which is given as 
\be
{\cal S}=-k_B \ln \bigg [1\pm exp(-2\pi r^2/{\lambda^2})\bigg ]
\ee
Now if we multiply the entropy $\cal S$ with temperature T we have energy,
which here appears as the interaction potentail energy.
\be
{\cal V}= {\cal S} T=-k_B T \ln \bigg [1\pm exp(-2\pi 
r^2/{\lambda^2})\bigg ]
\ee
This is exactly the interaction potential \cite{huang} used in our 
theory. 
\section{Behaviour of the wave function when $T\rightarrow\infty$ and
$r\rightarrow 0$}
The interaction potential Eq.(\ref{int}) in the limit of high T 
behaves\cite{arf} as
\be
\lim_{T\rightarrow \infty}
\tilde{\tilde v}(r,T)=-{T^{\frac{1}{4}}}{\delta(r)}
\ee
And our trial wave function Eq.(\ref{trial}) behaves\cite{arf} in the 
limit of high 
T as
\be
\lim_{T\rightarrow \infty}\chi(r,T)=c{T^{\frac{1}{4}}} r{\delta(r)}
\ee
where $c$ is a function of $\hbar$ and $m$ and independent of $r$ and  
T.
If we use the high T limit of interaction potential Eq.(B1) in the 1-D 
radial
Schr\"odinger Eq.(\ref{1d}), the above hight T limit of $\chi(r,T)$ 
[Eq.(B2)] does satisfy the
equation with an eigen energy  which is not bounded from below.
The eigen value diverges. This is exactly the behaviour
of the variational energy value[Eq.(\ref{energy})]  which also diverges in 
the high limit of T.
\end{document}